\documentclass[aps,pre,twocolumn,showpacs,superscriptaddress]{revtex4-1}

\usepackage{amsmath}
\usepackage{amssymb}
\usepackage{graphicx}
\usepackage{lineno}

\begin{document}

\title{Role of relaxation time scale in noisy signal transduction}

\author{Alok Kumar Maity}
\email{alokkumarmaity@yahoo.co.in}
\affiliation{Department of Chemistry,
University of Calcutta, 92 A P C Road, Kolkata 700 009, India}

\author{Pinaki Chaudhury}
\email{pinakc@rediffmail.com}
\affiliation{Department of Chemistry,
University of Calcutta, 92 A P C Road, Kolkata 700 009, India}

\author{Suman K Banik}
\email{Corresponding author; skbanik@jcbose.ac.in}
\affiliation{Department of Chemistry,
Bose Institute, 93/1 A P C Road, Kolkata 700 009, India}

\date{\today}

\begin{abstract}
Intracellular fluctuations, mainly triggered by gene expression, are an 
inevitable phenomenon observed in living cells. It influences generation 
of phenotypic diversity in genetically identical cells. Such variation of 
cellular components is beneficial in some contexts but detrimental in 
others. To quantify the fluctuations in a gene product, we undertake an
analytical scheme for studying few naturally abundant linear as well as 
branched chain network motifs. We solve the Langevin equations 
associated with each motif under the purview of linear noise approximation 
and quantify Fano factor and mutual information. Both quantifiable 
expressions exclusively depend on the relaxation time (decay rate 
constant) and steady state population of the network components. We 
investigate the effect of relaxation time constraints on Fano factor and 
mutual information to indentify a time scale domain where a network can 
recognize the fluctuations associated with the input signal more reliably. 
We also show how input population affects both quantities. We extend 
our calculation to long chain linear motif and show that with increasing 
chain length, the Fano factor value increases but the mutual information 
processing capability decreases. In this type of motif, the intermediate 
components are shown to act as a noise filter that tune up input fluctuations 
and maintain optimum fluctuations in the output. For branched chain motifs,
both quantities vary within a large scale due to their network architecture 
and facilitate survival of living system in diverse environmental conditions.
\end{abstract}


\maketitle


\section{Introduction}

Cell, building block of every biological system, is composed of some 
novel identical network motifs 
\cite{Alon2006, Alon2007, Barabasi2004, Milo2002, Tyson2003}. 
The ubiquitous examples of biological motifs are signal transduction network (STN) 
\cite{Marshall1995, Robison1997, Heinrich2002, Laub2007,Hynes2013}, 
gene transcription regulatory network (GTRN) 
\cite{Lee2002,ShenOrr2002, Blais2005, Boyer2005}, 
metabolic reaction network 
\cite{Jeong2000, Duarte2007, Forster2003} 
as well as protein-protein interaction network 
\cite{Han2004, Stelzl2005}. 
As cellular environment is stochastic in nature, it is interesting to investigate 
how these networks perform under fluctuating condition 
\cite{Kaern2005, Losick2008, Bruggeman2009, Kittisopikul2010, Rotem2010, 
Charlebois2011, Munsky2012}. The extent of performance of a network 
is measured by the response time, i.e., how fast the network output is 
changed with the fluctuating input stimuli \cite{Armitage1999}. If the 
network input-output relation follows a characteristic time scale then 
it can sense the extra cellular changes more precisely via network 
components. Consequently, some intracellular changes are ensured 
with the variation of input signal with few chemical modifications to 
optimize the environmental effect. One of the well studied STNs is 
mitogen activated protein kinase (MAPK) cascade, mostly observed 
in eukaryotic signaling pathway \cite{Marshall1995, Robison1997, Heinrich2002}. 
In MAPK cascade, external signal is processed through several steps 
via phosphorelay mechanism and fluctuations in signaling molecules 
get modified at every step of the cascade. This phenomenon has been 
verified by using both experimental and theoretical studies in early 
literature, where the authors have shown that output fluctuations increase 
in an integrated way with cascade length
\cite{Thattai2002, Hooshangi2005, Pedraza2005, TanaseNicola2006}. 
In the case of GTRN, the network is depicted by few nodes that represent 
regulatory genes and are connected with few edges. A simplest motif can
be constructed by considering two nodes representing two genes connected 
by a directed single edge. The edge signifies that product of one gene 
(transcription factor) regulates the other gene and the direction of the 
edge represents the mode of regulation. In such case, transcription factor 
may act as the signaling molecule and plays a pivotal role in maintaining the 
transcription rate of a target gene by controlling the appropriate time scale 
and the amount of transcriptional yield. GTRN motifs were initially identified 
in \textit{E. coli} where few network motifs are much more common compared 
to the other random motifs \cite{ShenOrr2002, Mangan2003}. Later, these 
common motifs have been also observed in several other prokaryotes 
as well as in eukaryotes. To understand cellular physiology, it is very much
essential to study GTRN motifs at the single cell level, because cell shows 
phenotypic heterogeneity in genetically identical system due to stochastic 
nature of these motifs. To this end, we have chosen few GTRN motifs to 
investigate the mode of functionality and their response under fluctuating
environment.

\begin{figure}[!t]
\begin{center}
\includegraphics[width=0.75\columnwidth,angle=0]{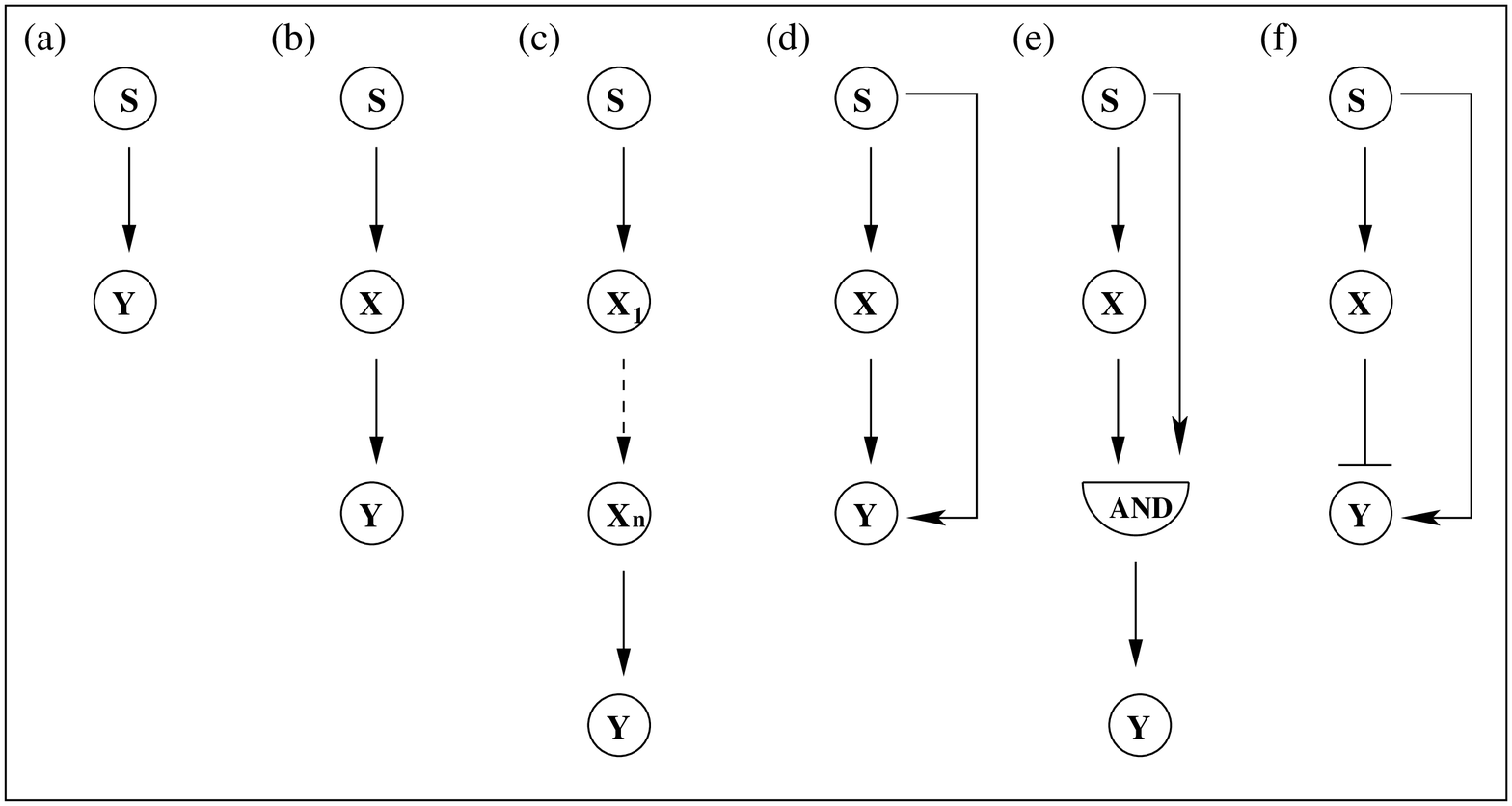}
\end{center}
\caption{Schematic presentation of different GTRN motifs 
(a) one step cascade (OSC),
(b) two step cascade (TSC),
(c) multi step cascade with $n$ number of intermediate nodes,
(d) OR coherent feed forward loop (OCFFL), 
(e) AND coherent feed forward loop (ACFFL) and
(f) incoherent feed forward loop (ICFFL).
}
\label{fig1}
\end{figure}

In the present communication, we focus on few naturally abundant 
GTRN network motifs. At first, we consider a simple linear one step cascade 
(OSC) and study the steady state dynamical behavior under stochastic 
framework. Two nodes and one edge are used to draw this motif, where 
each node represents a gene of a simple regulatory system (see Fig.~\ref{fig1}(a)). 
The edge indicates that one gene regulates the other via interaction of 
transcription factor with the promoter site of the target gene. Due to the direct
interaction of the two nodes, this motif can be considered as a direct pathway
for gene regulation or signal transmission. The next motif we undertake
is a linear two step cascade (TSC) obtained from the previous motif by 
inserting a new node (gene) in between the two nodes of the one step 
network (see Fig.~\ref{fig1}(b)). In this case, the target gene is indirectly 
regulated by the input (transcription factors) that act as a signal.

Using the OSC and the TSC network motifs, we then construct some common 
motifs of biological importance that belong to a group, i.e., feed forward loop 
(FFL) (see Fig.~\ref{fig1}(d-e)). We compose these motifs by lateral 
combination of two linear cascades (OSC and TSC). In FFL, a transcription 
factor regulates the target gene directly (via OSC) as well as indirectly (via TSC) 
\cite{ShenOrr2002, Mangan2003, Alon2006, Alon2007, Tsang2007}. In these 
motifs, two transcription factors are present and each of these can show either 
positive (activation) or negative (repression) effect on the target gene. Therefore, 
eight different types of FFL are possible considering both effects. Among all the
possible FFL, four of these are of coherent type and the remaining four are of 
incoherent type. The classification is done according to the sign of the overall 
regulatory motif, positive and negative sign for coherent and incoherent type, 
respectively. Experimentally, it has been shown that type-1 FFL has both 
coherent and incoherent nature and are ubiquitous. Due to this reason, we 
consider these two motifs in the present work. Type-1 coherent FFL 
has two sub-types depending upon the function of direct and indirect 
regulatory pathways on the promoter region of the target gene. When both 
transcription factors are required to express the target gene, the FFL motif
behaves as an AND like gate(see Fig.~\ref{fig1}(e)). On the other hand, when one
of the two transcription factors are sufficient to regulate the target gene, the
FFL motif behaves as a OR like gate (see Fig.~\ref{fig1}(d)). At this point, it is
important to mention that few theoretical studies under stochastic framework 
have been undertaken to understand the FFL motif
\cite{Ghosh2005,Hayot2005,Bruggeman2009,Murugan2012}.

We use a Gaussian model (see Models and Methods) to study the origin 
and consequence of 
stochasticity for all motifs considered in the present work. In all motifs, 
fluctuations are carried forward from one node to the next when 
signal is transduced along the direction of each edge. Thus, our main 
purpose here is the quantification of fluctuations in output signal for all 
motifs. Using an approximation technique (linear noise approximation), 
we solve all dynamical equations and calculate the Fano factor 
(variance/mean) \cite{Fano1947} expression by which we measure 
output fluctuations of each motif. We also try to understand the effect of 
relaxation time scale, i.e., lifetime of a network component on output 
fluctuations as it can provide knowledge about each and every step of 
fluctuation propagation through a cascade. In other words, relaxation 
time scale of each network component provides a way to measure the 
amplification or suppression of fluctuations in each step of signal 
propagation. We derive a time scale condition in which fluctuations in 
the input signal are filtered out by the intermediate component. Similarly,
conditions have been figured out when fluctuations are enhanced. We 
also examine the effect of copy numbers of input signal on output 
fluctuations and show that it plays a vital role under some specific 
conditions. As all cascades process information of the external signal, 
we investigate the reliability of information flow through each cascade by 
measuring the mutual information between the input signal and the 
output \cite{Borst1999, Mehta2009}.
We calculate the similar properties for FFL and identify different biological 
significance between two sub types of coherent motifs.

The rest of the paper is organized as follows. In the next section, we discuss
about the generic model and methods employed in the present work. In
Sec.~III, results of individual motifs have been discussed. The paper is 
concluded in Sec.~IV.


\section{Models and Methods}

To start with, we consider few network motifs that represent GTRN.
All the motifs that are taken into account in the present work are shown 
in Fig.~\ref{fig1} where each circle represents a node and a straight 
line with an arrowhead connecting two different nodes represents an 
edge. The direction of an arrowhead denotes flow of signal from one 
node to the next one. The simplest linear signal transduction motif is 
modeled by two nodes, where S component acts as an input signal node 
(transcription factor) which regulates the expression of a target gene Y 
(Fig.~\ref{fig1}(a)). The linear motif length is increased further by 
incorporating another node X in between the S and the Y nodes. In 
such a case, the S component regulates or transduces input signal 
to the Y component via the intermediate X (Fig.~\ref{fig1}(b)) 
\cite{Milo2002, Faisal2008, deRonde2012}. We also consider a long 
linear chain motif (Fig.~\ref{fig1}(c)) by integrating $n$ numbers of 
intermediate nodes within the simplest motif shown in Fig.~\ref{fig1}(a).

Next, we focus on few branched chain network motifs that are constructed 
by lateral combination of the first  two signal transduction motifs (one step 
and two step, Figs.~\ref{fig1}(a-b)) in different ways and are characterized 
as feed forward loop (FFL). Fig.~\ref{fig1}(d) represents coherent feed-forward 
loop of OR like gate (OCFFL), where the target gene Y is positively regulated 
by either the S or the X component, both acting as transcription factors. 
On the other hand, 
for coherent feed-forward loop of AND like gate (ACFFL), both S and X are 
essential to regulate the target gene Y positively (Fig.~\ref{fig1}(e)). In 
Fig.~\ref{fig1}(f), the transcription factor S positively regulates the production 
of gene Y in direct pathway but represses the gene regulation via the X 
mediated indirect pathway and this motif is known as incoherent feed-forward 
loop (ICFFL) \cite{Mangan2003, Alon2007, deRonde2012}.

All the biochemical network motifs considered in Fig.~\ref{fig1} consist of 
an input signal S and a output signal Y with an intermediate X except the 
long chain linear motif with $n$ numbers of intermediate components
(see Fig.~\ref{fig1}(c)). We describe the time dependent dynamics of the 
three chemical components by a set of generic coupled Langevin equations 
which may be of linear or non linear type depending on the kinetic schemes 
of a network motif considered in the present work
\begin{eqnarray}
\frac{ds}{dt} & = & f_s (s) - \tau_s^{-1} s + \xi_s (t), 
\label{eq1} \\
\frac{dx}{dt} & = & f_x (s,x) - \tau_x^{-1} x + \xi_x (t), 
\label{eq2} \\
\frac{dy}{dt} & = & f_y (s,x,y) - \tau_y^{-1} y + \xi_y (t).
\label{eq3}
\end{eqnarray}

\noindent
In Eqs.~(\ref{eq1}-\ref{eq3}), $f_i$ and $\tau^{-1}_i$ represent the functional 
form of synthesis and degradation (inverse of life time $\tau_i$) rate constants 
of the $i$-th ($i=s,x,y$) chemical component, respectively. Here, $s$, $x$ 
and $y$ stand for the chemical species S, X and Y, respectively. The noise 
terms $\xi_i$ is considered to be Gaussian white noise with zero mean. 
The noise strength or the variance associated with each noise term can
be written as $\langle | \xi_i |^2 \rangle$, quantified by the sum of 
production and decay rate. The cross-correlation between two noise terms
is zero as the two kinetics are uncorrelated with each other. Considering
that the copy number of each component is large at steady state, we study the
dynamics of each motif shown in Fig.~\ref{fig1}, at steady state. Since the 
relaxation time of each component is small compared to the coarse grained
(steady state) time scale, the Langevin equations we have adopted could 
satisfactorily explain the dynamics of each motif.

To solve the set of Langevin equations in a generalized way, we write these 
equations in the matrix form. To this end, we introduce two column vectors 
$z$ and $\xi$ where $z = (s,x,y)$ and $\xi = (\xi_s, \xi_x, \xi_y)$. Linearizing 
the Langevin equations around steady state and considering the change in 
copy number due to fluctuations of each species from steady state to be very 
small, one can write $\delta z(t) = z(t) - \langle z \rangle$, where 
$\langle z \rangle$ is the steady state value of $z$
\cite{Elf2003, Swain2004, Bialek2005, Kampen2005, deRonde2010}. 
Performing Fourier transformation of the linearized equation, we obtain
\begin{equation}
i \omega \delta \tilde{z} (\omega) = J_{z=\langle z \rangle}
\delta \tilde{z} (\omega) + \tilde{\xi} (\omega),
\label{eq4}
\end{equation}

\noindent
where $\tilde{z} (\omega)$ and $\tilde{\xi} (\omega)$ are Fourier transforms
of $z(t)$ and $\xi (t)$, respectively. $J$ is the Jacobian matrix evaluated at 
steady state. The diagonal elements of $J$ define the relaxation time of 
each component and the off-diagonal terms take care of interaction between 
the two components. The power spectra of the network components can be
derived using Eq.~(\ref{eq4})
\begin{equation}
{\cal S} (\omega) = \left [ i\omega I - J \right ]^{-1} H 
\left [ -i\omega I - J^T \right ]^{-1},
\label{eq5}
\end{equation}

\noindent
with $I$ being the identity matrix. Elements of $H$ stands for noise strength
and $J^T$ is the transpose of $J$. We perform the inverse Fourier transformation
of power spectra for every network component at steady state and evaluate 
the variance as well as covariance of the individual component and between 
two components, respectively. From the variance of output component, one 
can quantify the extent of fluctuations that are transduced by the final transcript 
of all the network motifs considered and the quantity of fluctuations can be 
defined in terms of Fano factor $\sigma_{y}^2/ \langle y \rangle$, ratio of 
variance and population of Y component \cite{Fano1947}. We also calculate 
mutual information ${\cal I}(s,y)$ between the input signal and the output using 
Shannon formalism to check the reliability of all network motifs 
\cite{Shannon1948, Cover1991, Borst1999}
\begin{equation}
{\cal I} (s,y) = \frac {1}{2} \log_{2} \left [1 + \frac {\sigma_{sy}^4}
{\sigma_{s}^2 \sigma_{y}^2 - \sigma_{sy}^4} \right ],
\label{eq6}
\end{equation}

\noindent
where $\sigma_{s}^2$ and $\sigma_{y}^2$ are the variance associated
with the S and the Y component, respectively, and the covariance between 
them is given by $\sigma_{sy}^2$.


\section{Results and discussions}

In the following subsections, we execute individual study of each network 
motif as well as perform a comparative study of all the motifs. From the 
Fano factor expression, we can discriminate the origin of fluctuations of a 
motif. We identify the network that faces maximum fluctuating environment 
under a definite condition and characterize the favorable circumstances in 
which it can transduce the information of the input signal more reliably.

\subsection{One step cascade}

As one step cascade (OSC) is the simplest unit of addressing network 
motifs, we initially start with this simple motif. In this motif, the input signal
S directly regulates the target gene Y. For the sake of simplicity, we 
consider that S is constitutively active and linearly regulates the target 
gene for the formation of Y. The stochastic reaction scheme in Langevin 
formalism is given by
\begin{eqnarray}
\frac{ds}{dt} & = & k_1 - \tau_s^{-1} s + \xi_s (t), 
\label{eq7} \\
\frac{dy}{dt} & = & k_3 s - \tau_y^{-1} y + \xi_y (t),
\label{eq8}
\end{eqnarray}

\noindent
where $k_1$ and $k_3$ are the synthesis rate for $S$ and $Y$, respectively.
The degradation rates for the same components are given by $\tau_s^{-1}$ 
and $\tau_y^{-1}$, respectively. $\xi_s (t)$ and $\xi_y (t)$ are Gaussian
white noise terms with zero mean 
$\langle \xi_s (t) \rangle = \langle \xi_y (t) \rangle = 0$. 
The respective noise strengths are given by
$\langle \xi_s (t) \xi_s (t') \rangle = 2 \tau_s^{-1} \langle s \rangle
\delta (t-t')$ and 
$\langle \xi_y (t) \xi_y (t') \rangle = 2 \tau_y^{-1} \langle y \rangle
\delta (t-t')$, respectively. In addition, both noise processes are
uncorrelated, 
$\langle \xi_s (t) \xi_y (t') \rangle = \langle \xi_y (t) \xi_s (t') \rangle
= 0$.
We solve Eqs.~(\ref{eq7}-\ref{eq8}) by the above mentioned procedure 
and calculate the variance associated with the output Y and co-variance 
between the input signal S and the output Y as 
\cite{Elf2003, Swain2004, Bialek2005, Kampen2005, deRonde2010}
\begin{equation}
\sigma_y^2 = \langle y \rangle
+ \frac{\tau_y^{-1} \langle y \rangle^2
}{(\tau_s^{-1}+\tau_y^{-1})\langle s \rangle} ,
\sigma_{sy}^2 = 
\frac{\tau_y^{-1} \langle y \rangle}{\tau_s^{-1}+\tau_y^{-1}} .
\label{eq9}
\end{equation}

\noindent
In the above expression of variance $\sigma_{y}^2$, the first part on the right 
hand side arises due to intrinsic fluctuations in Y and the second part is 
responsible for extrinsic fluctuations, incorporated into the output Y through 
the input S during gene regulation. The total variance in this motif is expressed 
in terms of output variance $\sigma_{y}^2$ that follows spectral addition rule, 
sum of external and internal fluctuations 
\cite{Elowitz2002, Swain2002, Raser2004, Paulsson2004, TanaseNicola2006, Hilfinger2011}. 
At this point, it is important to mention that two fluctuating terms originating 
from different sources have been also calculated for the oscillatory system 
\cite{Scott2006}. 
In this study, we are interested in Fano factor as well as in information 
propagation through the cascade with the variation of system's relaxation 
times as well as steady state population $\langle s \rangle$ of the 
input component S, as both $\sigma_y^2$ and $\sigma_{sy}^2$ depend 
only on the time scale when the steady state value of both components 
$\langle s \rangle$ and $\langle y \rangle$ are kept fixed followed by a 
constant $k_1/ \tau_{s}^{-1}$ and $k_3/ \tau_{y}^{-1}$ ratio. 
Similarly, for constant relaxation times, both expressions vary with the steady 
state populations $\langle s \rangle$ and $\langle y \rangle$. In this motif, 
we have two relaxation time scales $\tau_s$ (input) and $\tau_y$ (output).
These two time scales lead to three possible limiting conditions for which
we get three different modified expressions for Fano factor 
($\sigma_{y}^2/ \langle y \rangle$) as well as for co-variance (see
Table~\ref{table1}).


\begin{table}[!t]
\caption{\label{table1} 
Modified form of the analytical expression given by Eq.~(\ref{eq9}) for OSC motif.
Fano factor ($\sigma_{y}^2/ \langle y \rangle$) and co-variance ($\sigma_{sy}^2$) 
for different relaxation time limit are shown in this table.
}
\begin{ruledtabular}
\begin{tabular}{cccc}
& $\tau_s \gg \tau_y$ & $\tau_s \approx \tau_y$ & $\tau_s \ll \tau_y$ \\
\hline
Fano factor & $1+\frac{\langle y \rangle}{\langle s \rangle}$ 
& $1+0.5 \frac{\langle y \rangle}{\langle s \rangle}$
& $1+\frac{\tau_s \langle y \rangle}{\tau_y \langle s \rangle}$ \\
$\sigma_{sy}^2$ & $\langle y \rangle$ & $0.5 \langle y \rangle$
& $\frac{\tau_s}{\tau_y} \langle y \rangle$ \\
\end{tabular}
\end{ruledtabular}
\end{table}

In Table~\ref{table1}, Fano factor and $\sigma_{sy}^2$ values are maximum 
at the time limit $\tau_s\gg \tau_y$ whereas, both are minimum at the time limit 
$\tau_s\ll \tau_y$. These results reveal that the effect of input fluctuations into 
the output fluctuations gets maximized if the input signal relaxes at a much 
slower rate compared to the output signal ($\tau_{s}^{-1}\ll \tau_{y}^{-1}$) and 
will be minimized for faster input relaxation rate compared to the output one 
($\tau_{s}^{-1}\gg \tau_{y}^{-1}$). However, the output signal faces an intermediate 
level of fluctuations when both signals have comparable relaxation rate 
($\tau_{s}^{-1}\approx \tau_{y}^{-1}$).

\begin{figure}[!t]
\begin{center}
\includegraphics[width=0.75\columnwidth,angle=0]{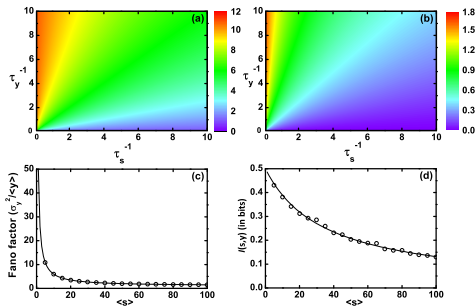}
\end{center}
\caption{(Color online) The OSC. 
(a, b) Two dimensional maps of Fano 
factor and mutual information ${\cal I}(s,y)$, respectively, as a function 
of two relaxation rate constants $\tau_{y}^{-1}$ and $\tau_{s}^{-1}$ for 
the ratio $k_1/\tau_{s}^{-1} = k_3/\tau_{y}^{-1} = 10$.
(c) Fano factor and (d) mutual information ${\cal I}(s,y)$ profiles as a
function of mean input signal level $\langle s \rangle$. 
Parameters used are $\tau_{s}^{-1} = \tau_{y}^{-1} = 1.0$ and 
$k_3 = 100/k_1$. 
The symbols are generated using stochastic simulation algorithm 
\cite{Gillespie1976,Gillespie1977} and the lines are due to theoretical 
calculation.
}
\label{fig2}
\end{figure}

For faster fluctuations in the input component ($\tau_s \ll \tau_y$), the target 
gene cannot sense the rapid concentration changes of the input signal and 
shows an average response. In such a case, external fluctuations have no 
significant contribution in the fluctuations of output and consequently, 
suppression of the output fluctuations is executed and minimum Fano factor 
value is obtained. In this connection, it is important to mention that for very 
large $\tau_y$, the ratio $\tau_s / \tau_y$ is very low ($\ll 1$) and contribution 
of extrinsic fluctuations becomes insignificant in the total output fluctuations. 
Therefore, output fluctuations only depend on the mean steady state value 
of the target gene and the network motif follows a Poisson statistic, i.e., 
behaves like a simple birth-death process (Fano factor $\sigma_{y}^2/ \langle y \rangle=1$).         
However, the target gene successfully characterizes the concentration 
change of signaling molecule for slower input fluctuations ($\tau_s\gg \tau_y$). 
In this time scale, the OSC motif transduces extracellular or upstream 
signal reliably and provides an exact response with the achievement of 
maximum Fano factor. When both time scales are approximately equal 
($\tau_s\approx \tau_y$), extrinsic fluctuations get partially incorporated 
into the total fluctuations and give an intermediate Fano factor value which
is in between the two extreme cases, slowest and fastest input fluctuations. 
It seems apparent that the motif can sense external fluctuations with a greater 
extent in the nearly equal relaxation time scales but our result does not show
that. As both reactions are stochastic in nature, they have probabilistic 
character that executes two chance factors. In one situation, the target gene 
can properly characterize the input fluctuations and in the other situation, it fails 
to characterize the same. This results into a statistical weightage value of
0.5 in the contribution of extrinsic fluctuations in the Fano factor expression. 
A similar kind of time scale effect is also shown by the co-variance expression 
which governs the mutual information transduction. In Figs.~\ref{fig2}(a-b), 
we show surface plots of Fano factor and mutual information, respectively, 
as a function of two relaxation rate constants, $\tau_{y}^{-1}$ and 
$\tau_{s}^{-1}$, where we maintain the steady state population of both 
components by a constant parameters ratio 
$k_1/ \tau_{s}^{-1}=k_3/ \tau_{3}^{-1}=10$. Fig.~\ref{fig2}(a) shows 
that the maximum Fano factor value is attained by the motif only at very low 
input relaxation rate constant compared to the output one. Along diagonal 
axis both rate constants are approximately equal. Hence, the magnitude of 
Fano factor is within an intermediate range. The minimum level of Fano 
factor value is observed at very high input rate constant compared to the output one. 
In Fig.~\ref{fig2}(b), the 2d-plot of mutual information also varies in a similar 
fashion as the Fano factor plot. As the OSC motif performs under the definite 
input fluctuations, the information transduction capability of the motif is mainly 
characterized on the basis of input-output relaxation time scales.  As a result, 
the motif can transude the input information more reliably at faster relaxation 
time scale of the output component among all the relaxation time scales of 
output component (see the three limiting conditions in Table~\ref{table1}).


For the three relaxation time limiting cases given in Table~\ref{table1},
the Fano factor expression also depends on the steady state population 
of the network components. However, from the Fano factor expression it is
clear that extrinsic fluctuations can contribute an appreciable amount in the 
total output noise if the steady state population of output component is 
much higher than the input one. This means that for a highly populated 
input signal, the fluctuations in the input do not have significant effect 
on the regulation of the target gene. Consequently, the regulated gene works 
under an apparently constant level of the input signal. Therefore, the 
motif shows low level of Fano factor value even if it belongs to the 
relaxation time scale limit where maximum input fluctuations are transduced. 
In Fig.~\ref{fig2}(c), we show the Fano factor associated with the variation of 
population of the signaling molecule S. In this plot, a sharp exponential 
decay of the Fano factor is observed as the steady state population of the
input signal increases. In Fig.~\ref{fig2}(d), mutual information 
${\cal I}(s,y)$ is plotted as a function of the input signal. In Figs.~\ref{fig2}(c-d), 
population level of the input signal is systematically increased by increasing 
the synthesis rate constant $k_1$ whereas, the steady state population
of the output Y is kept fixed by simultaneous change of the synthesis rate 
constant $k_3$, followed by a mathematical relation $k_3=100/k_1$. The 
rest of the parameter values used are $\tau_1^{-1}=\tau_3^{-1}=1.0$. In the 
mutual information plot (Fig.~\ref{fig2}(d)), sharp exponential decay is not 
observed as in Fig.~\ref{fig2}(c). This happens due to the absence of explicit 
dependence of signaling molecule in the expression of co-variance (see 
Eq.~(\ref{eq9})) and for keeping $\langle y \rangle$ fixed. In addition, in the 
expression of mutual information (see Eq.~(\ref{eq6})), the input signal has 
a predominant effect due to $\sigma^2_s$ that overcomes the decreasing 
effect of $\sigma^2_y$ associated with Y.

Based on the analysis provided in the aforesaid discussion, one can compare 
the OSC motif with the well known motif of gene transcription and translation 
machinery. Here, mRNA, the product of transcription can be considered as
signaling component S generated from a fully induced or a constitutive promoter
at a constant rate. Similarly, protein, the product of translation plays the role
of component Y translated from mRNA. If one does not take into account the 
genetic switching steps (on/off state of a promoter) then one can easily compare
the gene regulation network with the OSC motif where mRNA (S) and protein
(Y) represent the input and the output component, respectively. From the
conditions presented in Table~\ref{table1}, one can conclude that the gene
regulation motif can only attain a low level of fluctuations in the protein level 
through high population and relaxation rate constant of the input component 
compared to the output one. Thus, fluctuations in the protein Y, the gene product,
are modulated via kinetic parameter as well as deterministic population of mRNA 
which acts as signaling component S. Protein molecule shows minimum 
fluctuations under the constraints of large number of mRNA with very low 
average lifetime $\tau_s$. These phenomena have been verified extensively
via experimental and theoretical studies
\cite{Thattai2001, Elowitz2002, Ozbudak2002, Blake2003, Fraser2004, Raser2004, Thattai2004, Kaern2005, Becskei2005, Raj2008}. 
Similarly, an extensive study on noise propagation in eukaryotic gene 
expression has been accomplished using data from two high-throughput 
experiments where Fraser et al \cite{Fraser2004} have observed that 
the production of essential and complex-forming proteins implicate low level 
of fluctuations compared to other proteins and this low level of fluctuations 
is attained via high transcription rate and low translation efficiency. Due to a
high transcription rate, a large amount of mRNA is generated. While doing the
analysis, they have used the definition of the translation efficiency, ratio of 
protein synthesis and mRNA decay rate constant. Translation efficiency can 
be minimized for very high decay (relaxation) rate constant of mRNA molecule,
i.e., very short lifetime under a constant protein production rate. From our 
calculation, we also get an equivalent result that successfully explicates their 
noble findings.

\subsection{Two step cascade}

In two step cascade (TSC)(see Fig.~\ref{fig1}(b)), the output component Y is 
indirectly regulated by the input component S via an intermediate component
X. As the dynamical equation for the input signal S is same as in the previous 
motif (Eq.~(\ref{eq7})), we do not write it here explicitly. The Langevin equations 
for the remaining two components X and Y are given as
\begin{eqnarray}
\frac{dx}{dt} & = & k_2 s - \tau_x^{-1} x + \xi_x (t), 
\label{eq10} \\
\frac{dy}{dt} & = & k_3 x - \tau_y^{-1} y + \xi_y (t).
\label{eq11}
\end{eqnarray}


\begin{table*}[!t]
\caption{\label{table2} 
Modified forms of the analytical solution (Eq.~(\ref{eq12})) of TSC motif.
Fano factor ($\sigma_{y}^2/\langle y \rangle$) and co-variance
($\sigma_{sy}^2$) at different relaxation time limits are shown
with $\rho=\tau_s/(\tau_s+\tau_y) \leqslant 1$.
}
\begin{ruledtabular}
\begin{tabular}{ccccc}
& &$\tau_x \gg \tau_y$ & $\tau_x \approx \tau_y$ & $\tau_x \ll \tau_y$ \\
\hline
$\tau_s \gg \tau_x$ & Fano factor & $1+\frac{\langle y \rangle}{\langle x \rangle}
+\frac{\langle y \rangle}{\langle s \rangle}$ 
& $1+0.5 \frac{\langle y \rangle}{\langle x \rangle}
+\frac{\langle y \rangle}{\langle s \rangle}$
& $1+\frac{\tau_x \langle y \rangle}{\tau_y \langle x \rangle}
+\rho \frac{\langle y \rangle}{\langle s \rangle}$ \\
&$\sigma_{sy}^2$ & $\langle y \rangle$ & $ \langle y \rangle$
& $\rho \langle y \rangle$ \\
$\tau_s \approx \tau_x$ & Fano factor & $1+\frac{\langle y \rangle}{\langle x \rangle}
+0.5\frac{\langle y \rangle}{\langle s \rangle}$ 
& $1+0.5 \frac{\langle y \rangle}{\langle x \rangle}
+\frac{3 \langle y \rangle}{ 8 \langle s \rangle}$
& $1+\frac{\tau_x \langle y \rangle}{\tau_y \langle x \rangle}
+\frac{\tau_x \langle y \rangle}{\tau_y \langle s \rangle}$ \\
&$\sigma_{sy}^2$ & $0.5 \langle y \rangle$ & $ 0.25 \langle y \rangle$
& $0.5 \frac{\tau_s \langle y \rangle}{\tau_y}$ \\
$\tau_s \ll \tau_x$ & Fano factor & $1+\frac{\langle y \rangle}{\langle x \rangle}
+\frac{\tau_s \langle y \rangle}{\tau_x \langle s \rangle}$ 
& $1+0.5 \frac{\langle y \rangle}{\langle x \rangle}
+0.5 \frac{\tau_s \langle y \rangle}{\tau_x \langle s \rangle}$
& $1+\frac{\tau_x \langle y \rangle}{\tau_y \langle x \rangle}
+\frac{\tau_s \langle y \rangle}{\tau_y \langle s \rangle}$ \\
&$\sigma_{sy}^2$ & $\frac{\tau_s \rho \langle y \rangle}{\tau_x}$ & 
$ \frac{\tau_s^2 \langle y \rangle}{\tau_x \tau_y}$
& $\frac{\tau_s^2 \langle y \rangle}{\tau_x \tau_y}$ \\
\end{tabular}
\end{ruledtabular}
\end{table*}

\noindent
In the above equations, $k_2$ and $k_3$ are the synthesis rate constants 
of X and Y component, respectively. $\tau_i^{-1}$ and $\xi_i(t)$ ($i=x,y$) are  
decay rate constants and Langevin force terms of the respective component. 
As in the OSC, the three noise terms ($\xi_s$, $\xi_x$ and $\xi_y$) here are 
Gaussian white type with similar noise properties. Solving the Langevin
equations in a similar manner 
\cite{Elf2003, Swain2004, Bialek2005, Kampen2005, deRonde2010},
we get the expression of variance and co-variance of the output component Y
\begin{eqnarray}
\sigma_y^2 &=& \langle y \rangle
+ \frac{\tau_y^{-1} \langle y \rangle^2
}{(\tau_x^{-1}+\tau_y^{-1})\langle x \rangle} \nonumber\\
&& + \frac{\tau_x^{-1}\tau_y^{-1} (\tau_s^{-1}+\tau_x^{-1}+\tau_y^{-1}) \langle y \rangle^2
}{(\tau_s^{-1}+\tau_x^{-1})(\tau_x^{-1}+\tau_y^{-1})(\tau_s^{-1}+\tau_y^{-1})\langle s \rangle}, \nonumber\\
\sigma_{sy}^2 &=& 
\frac{\tau_x^{-1}\tau_y^{-1} \langle y \rangle}{(\tau_s^{-1}+\tau_x^{-1})(\tau_s^{-1}+\tau_y^{-1})} .
\label{eq12}
\end{eqnarray}

\noindent
The first term on the right hand side of the variance $\sigma_y^2$ reveals 
the intrinsic fluctuations in the output component Y. The second and the third 
terms of the expression originate due to the fluctuations in the X and the S 
component, respectively. Compared to the OSC motif, an extra noise term 
appears in the variance which originates due to the addition of intermediate 
component X. Hence, the magnitude of total output fluctuations in TSC 
becomes higher compared to the OSC. If one inserts a new intermediate 
component into the TSC motif, the output fluctuations will increase further. This 
indicates that the output fluctuations are increased with the augmentation of 
cascade length. Such fluctuations integration character in each step of a 
cascade has been verified earlier both experimentally and theoretically 
\cite{Thattai2002, Hooshangi2005, Pedraza2005, TanaseNicola2006}. 
In spite of these fluctuations enhancement property of long chain cascade 
networks, some long chain network motifs like MAPK signaling pathways 
as well as GTRNs are identified in living systems, where external signal gets
transduced with great accuracy. This is an unusual but interesting aspect
of living beings that promotes an extra curiosity to study signaling pathways
to understand the execution of high precision signal transduction in highly
fluctuating environment. While considering the dynamics of TSC, we try to
decipher the criteria that leads to the understanding of the formation of
final output Y under the condition of optimum fluctuations as per the system's
permissibility. In Eq.~(\ref{eq12}), both expressions depend on the three
relaxation rate constants $\tau_s^{-1}$, $\tau_x^{-1}$ and $\tau_y^{-1}$ as 
well as on the steady state population level of the network components 
($ \langle s \rangle$, $ \langle x \rangle$ and $ \langle y \rangle$). 
Possible combinations of three relaxation times ($\tau_s$, $\tau_x$ and 
$\tau_y$) leads to nine different modified expressions of Fano factor 
($\sigma_{y}^2/\langle y \rangle$) and co-variance ($\sigma_{sy}^2$),
shown in Table~\ref{table2}.

\begin{figure}[!t]
\begin{center}
\includegraphics[width=0.75\columnwidth,angle=0]{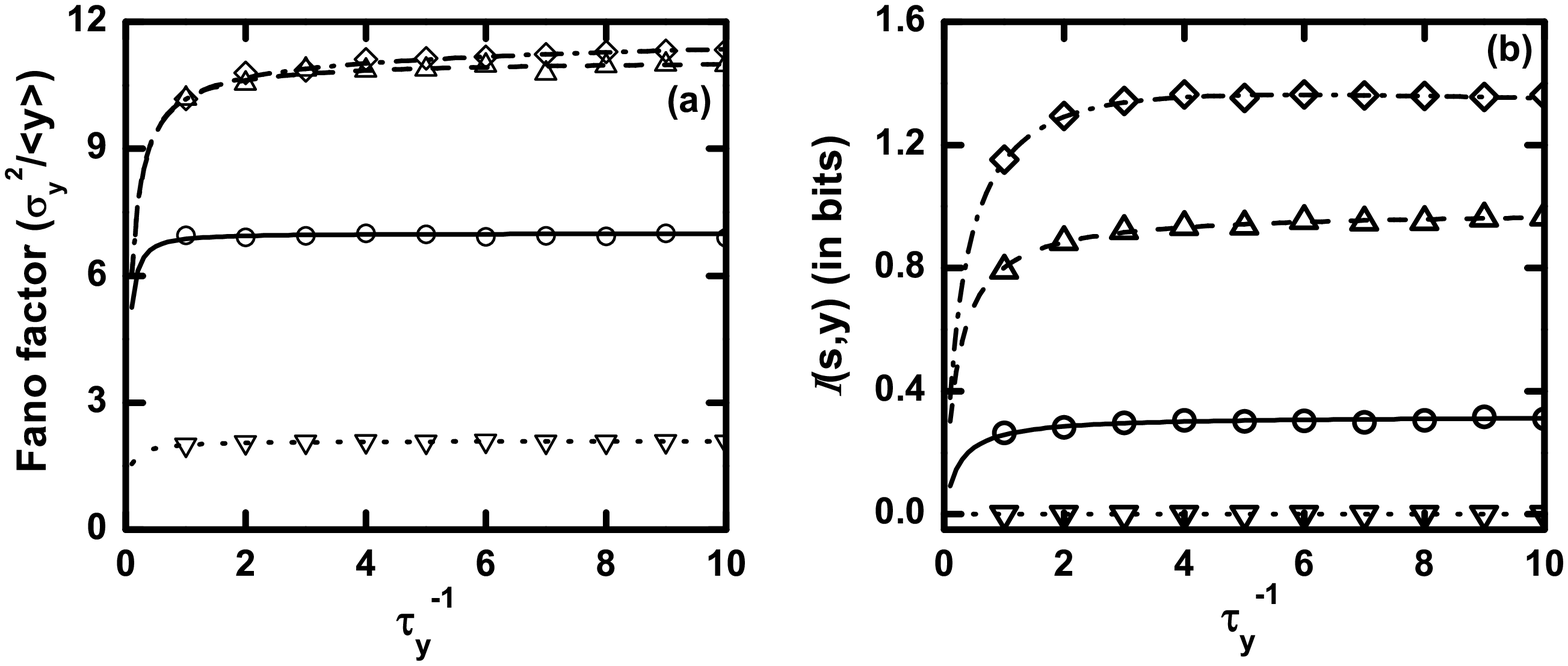}
\end{center}
\caption{The TSC. 
(a) Fano factor and (b) mutual information ${\cal I}(s,y)$ profiles 
as a function of relaxation rate constant $\tau_y^{-1}$ of Y component.
$k_1/\tau_{s}^{-1} = k_2/\tau_{x}^{-1} = 10$ and $k_3/\tau_{y}^{-1} = 1.0$ 
are maintained throughout these plots, so that steady state population 
of all the components remains unaltered. 
In both plots, for solid (with open circles), dashed (with open upward triangle), 
dotted (with open downward triangle) and dash dotted (with open diamond) 
lines we have used the relations
$\tau_{s}^{-1} = \tau_{x}^{-1} = 0.1$, 
$\tau_{s}^{-1} = \tau_{x}^{-1}/10 = 0.1$, 
$\tau_{s}^{-1}/100 = \tau_{x}^{-1} = 0.1$ and 
$\tau_{s}^{-1} = \tau_{x}^{-1}/100 = 0.1$, 
respectively.
The symbols are generated using stochastic simulation algorithm 
\cite{Gillespie1976,Gillespie1977} and the lines are due to theoretical 
calculation.
}
\label{fig3}
\end{figure}

From the modified expressions given in Table~\ref{table2} it is easy to identify
the maximum, intermediate and minimum value of Fano factor and co-variance
under the condition $\tau_s \gg \tau_x \gg \tau_y$, 
$\tau_s \approx \tau_x \approx \tau_y$ and $\tau_s \ll \tau_x \ll \tau_y$, respectively. 
In all the expressions, effect of the population level of the input and the intermediate 
component on the output is clearly visible. As our main focus in the present study is 
to characterize the effect of relaxation time scales in terms of Fano factor and 
co-variance, we fix the steady state population of all the network components using
the relations $k_1/ \tau_s^{-1}=k_2/ \tau_x^{-1}=10$ and $k_3/ \tau_y^{-1}=1.0$.
In Figs.~\ref{fig3}, we show Fano factor and mutual information for the TSC motif 
as a function of output relaxation rate constant $\tau_y^{-1}$ for different values of
$\tau_s^{-1}$ and $\tau_x^{-1}$.


From the plot  shown in Fig.~\ref{fig3}(a), it is clear that Fano factor value 
increases with the increment of output relaxation rate constant $\tau_y^{-1}$ 
for all sets of parameter values. As the output component Y can track the 
comparably slower fluctuations of upstream input signal efficiently, 
fluctuations of the output component increases proportionately with its 
relaxation rate constant. The additive nature of fluctuations assists to amplify 
the Fano Factor value through total output as a function of $\tau_y^{-1}$ 
and the trend is persistent for all the four sets of parameter values used to
draw Fig.~\ref{fig3}(a). Among the four sets of parameters, higher Fano 
factor values are achieved by two sets of parameters. From the parameter
sets $\tau_s^{-1}=\tau_x^{-1}/10=0.1$ and $\tau_s^{-1}=\tau_x^{-1}/100=0.1$,
it is evident that the intermediate component X has faster fluctuations rate
than the input component S ($\tau_x^{-1} \gg \tau_s^{-1}$). As a result, it 
becomes possible for X to characterize the fluctuations in S. At the equal 
relaxation time scale limits of S and X ($\tau_s^{-1}=\tau_x^{-1}=0.1$), 
we get an intermediate Fano factor profile and a minimum profile of the same 
is obtained for $\tau_s^{-1}/100=\tau_x^{-1}=0.1$. This happens as the S
component fluctuates in a faster time scale compared to the X component 
which, in turn, forbids X to differentiate the concentration change in S. As a
result, X is unable to carry forward the variation in signal due to S to the Y
component of the TSC motif. In this relaxation time scale limit, the intermediate 
X acts as a low pass filter that passes only low-frequency input signal. For this 
reason, highly fluctuating (high-frequency) input signal is impeded by the 
intermediate component that inhibits fluctuations propagation through the 
motif.

In Fig.~\ref{fig3}(b), mutual information ${\cal I}(s,y)$ flow varies with the 
relaxation rate of Y and we observe that TSC motif transduces information 
more reliably at faster fluctuations in Y due to proper characterization of 
the input signal variation and is applicable for different choice of parameters 
sets. For the four sets of parameters, mutual information follows the increasing 
trend of Fano factor but, it is important to mention that for 
$\tau_s^{-1}/100=\tau_x^{-1}=0.1$, ${\cal I}(s,y)$ value is almost zero. This
happens due to slow fluctuations in X compared to S which impedes the
flow of information due to input signal into the downstream component. As a 
result, the intermediate component cannot discriminate the variation of 
external signal and gives a constant response to the change in the environment.
Although the Fano factor is minimum at this time frame, the TSC motif is
unable to adapt due to the approximately zero mutual information processing 
ability. Hence, the network motif opts for a relaxation time scale where both
fluctuations and mutual information processing capacity adopts an optimum
value.

Next, we extend our calculation for a generalized motif of linear long chain 
cascade (see Fig.~\ref{fig1}(c)) considering $n$ numbers of different intermediate 
components (X$_1$, X$_2$, É.., X$_n$) that are present in between the 
input and the output component. Using the concept of TSC motif, we obtain 
the expressions of Fano factor for three limiting relaxation rate conditions as
\begin{eqnarray}
&& 1 + \frac{\langle y \rangle}{\langle x_n \rangle} + 
\cdots + \frac{\langle y \rangle}{\langle x_1 \rangle} + 
\frac{\langle y \rangle}{\langle s \rangle}  \nonumber \\ 
&& \mbox{ for } \tau_y \gg \tau_{x_n}\gg \cdots \gg \tau_{x_1} \gg \tau_s, \nonumber \\
&& 1 + c_n \frac{\langle y \rangle}{\langle x_n \rangle} + 
\cdots + c_1 \frac{\langle y \rangle}{\langle x_1 \rangle}
+ c_s \frac{\langle y \rangle}{\langle s \rangle}  \nonumber \\ 
&& \mbox{ for } \tau_y \approx \tau_{x_n} \approx \cdots \approx \tau_{x_1} \approx \tau_s,
\nonumber \\
&& 1 + \frac{\tau_{x_n} \langle y \rangle}{\tau_y \langle x_n \rangle} + 
\cdots + \frac{\tau_{x_1} \langle y \rangle}{\tau_y\langle x_1 \rangle} +
\frac{\tau_s \langle y \rangle}{\tau_y \langle s \rangle}  \nonumber \\
&& \mbox{ for } \tau_y \ll \tau_{x_n} \ll \cdots \ll \tau_{x_1} \ll \tau_s. \nonumber
\end{eqnarray}

\noindent
where $c_1, \cdots, c_n \leqslant 0.5$, $c_s \leqslant 0.5$ and $\tau_{x_1}, 
\tau_{x_2}, \cdots, \tau_{x_n}$ are the relaxation time of the X$_1$, X$_2$, $\cdots$, 
X$_n$ component, respectively. The expressions given above are three 
simplified forms of Fano factor expressions for linear long chain signal 
transduction motif or GTRN cascade with $n$ number of different intermediate 
components. The main motivation to evaluate these simplified forms is that using
the expressions, one can easily get a gross quantitative idea about the output 
fluctuations of any linear chain network cascade with large number of intermediate
components. If one knows the steady state population level of the network 
components as well as lifetime of the same then only using those parameters
one can calculate Fano factor quantity that will provide a hint about the networkÕs 
fluctuations. This is the main advantage of the present formalism that makes 
easier the study of stochastic features of several unexplored systems.

In the aforesaid discussion, we have searched for the effect of relaxation time scale 
on signal transduction machinery through linear type of GTRN motifs. These
results suggest us to extend our analysis to motifs having branched pathway.
The branched pathway motifs are generated with the help of lateral combinations 
of more than one linear pathway motifs. Therefore, our next objective is to 
investigate the relaxation time scale effect on a family of branched network such 
as feed forward loop (FFL) \cite{ShenOrr2002, Mangan2003, Alon2007, Tsang2007}. 
FFL appears more frequently in gene networks of \textit{E. coli}, \textit{S. cerevisiae}
and other living organisms. It consists of three genes which are characterized by 
three different transcription factors S, X and Y where X regulates Y and S regulates 
both X and Y. Thus, S directly as well as indirectly, via X, regulates Y leading to
positive (activation) or negative (repression) transcription interaction. Sign of the 
direct regulation path is equal to the indirect path for coherent feed forward loop 
(CFFL) but opposite sign of two regulatory pathways is the basis for incoherent 
feed forward loop (ICFFL). Although eight FFL are possible from structural 
configuration, we study only two most abundant network motifs, type-1 CFFL and 
ICFFL. The FFL networks have two input signals, one signal induces the S encoded 
transcription factor gene and the other induces the X encoded gene. For the sake of 
simplicity, we investigate these motifs under the effect of constant input signal. In 
the following subsections, we discuss the role of different time scales on the two
FFL motifs.

\subsection{Coherent feed forward loop}

In type-1 CFFL motif, both S and X either positively or negatively regulate the 
promoter of the target gene Y.  The expression level of output Y is controlled 
by the concentration of two upstream transcription factors. If both S and X are 
required to control the production of Y, the motif behaves as AND like logic 
gate otherwise it behaves as an OR like logic gate. Keeping this in mind, we 
examine these two architectures and investigate which one of these two faces 
maximum fluctuations in a noisy environment.

\subsubsection{OR like gate}

In OR like CFFL (OCFFL) motif (Fig.~\ref{fig1}(d)), any one of the two upstream
components (S and X) is sufficient to control the expression of Y. Hence, in the
dynamical equation of Y, we introduce two different synthesis rate constants,
$k_3^{\prime}$ and $k_3$ for the direct and the indirect regulation pathways, 
respectively. These two rate constants give freedom to tune up the extent of 
interaction amid the transcription factor and the target gene in an independent 
way. The first term defines the extent of interaction to express Y via direct pathway 
and the second one is responsible for the indirect pathway. From the logical 
condition, either of the two interactions of this motif is predominant over the other, 
i.e., if the direct regulation is stronger then the indirect regulation must be weaker, 
or vice versa. When the direct regulation path is more prominent, the motif 
behaves like the OSC but if the indirect path plays a pivotal role, then it performs 
as the TSC. Keeping this in mind, we write the stochastic dynamical equations as
\begin{eqnarray}
\frac{dx}{dt} & = & k_2 s - \tau_x^{-1} x + \xi_x (t), 
\label{eq13} \\
\frac{dy}{dt} & = & k_{3}^{\prime} s + k_3 x - \tau_y^{-1} y + \xi_y (t).
\label{eq14}
\end{eqnarray}


\begin{table*}[!t]
\caption{\label{table3} 
Modified forms of the analytical solution (Eq.~(\ref{eq15})) of OCFFL motif.
Fano factor ($\sigma_{y}^2/\langle y \rangle$) and co-variance ($\sigma_{sy}^2$)
at different relaxation time limits are shown where 
$\rho=\tau_s/(\tau_s+\tau_y) \leqslant 1$,
$\alpha=k_2/\tau_x^{-1}$, $\beta=k_3/\tau_y^{-1}$, 
$\gamma=k_3^{\prime}/\tau_y^{-1}$ and
$p={\alpha}^2 {\beta}^2+2\alpha \beta \gamma$.
}
\begin{ruledtabular}
\begin{tabular}{ccccc}
& &$\tau_x \gg \tau_y$ & $\tau_x \approx \tau_y$ & $\tau_x \ll \tau_y$ \\
\hline
$\tau_s \gg \tau_x$ & Fano factor & $1+\frac{{\beta}^2 \langle x \rangle}{\langle y \rangle}
+\frac{({\gamma}^2 + p) \langle s \rangle}{\langle y \rangle}$ 
& $1+0.5 \frac{{\beta}^2 \langle x \rangle}{\langle y \rangle}
+\frac{({\gamma}^2 + p) \langle s \rangle}{\langle y \rangle}$
& $1+\frac{{\beta}^2 \tau_x \langle x \rangle}{\tau_y \langle y \rangle}
+\rho \frac{({\gamma}^2 + p) \langle s \rangle}{\langle y \rangle}$ \\
&$\sigma_{sy}^2$ & $\langle y \rangle - \frac{\tau_x \beta \langle x \rangle}{\tau_s}$ 
& $ \langle y \rangle- \frac{\tau_x \beta \langle x \rangle}{\tau_s}$
& $\rho (\langle y \rangle- \frac{\tau_x \beta \langle x \rangle}{\tau_s})$ \\
$\tau_s \approx \tau_x$ & Fano factor & $1+\frac{{\beta}^2 \langle x \rangle}{\langle y \rangle}
+\frac{({\gamma}^2 + 0.5p) \langle s \rangle}{\langle y \rangle}$ 
& $1+0.5 \frac{{\beta}^2 \langle x \rangle}{\langle y \rangle}
+\frac{({0.5 \gamma}^2 + \frac{3 p}{8}) \langle s \rangle}{\langle y \rangle}$
& $1+\frac{{\beta}^2 \tau_x \langle x \rangle}{\tau_y \langle y \rangle}
+\frac{\tau_s ({\gamma}^2 + p) \langle s \rangle}{\tau_y \langle y \rangle}$ \\
&$\sigma_{sy}^2$ & $\langle y \rangle - 0.5 \beta \langle x \rangle$ 
& $ 0.5 \langle y \rangle- 0.25 \beta \langle x \rangle$
& $\frac{\tau_s (\langle y \rangle- 0.5 \beta \langle x \rangle)}{\tau_y}$ \\
$\tau_s \ll \tau_x$ & Fano factor & $1+\frac{{\beta}^2 \langle x \rangle}{\langle y \rangle}
+\frac{(\rho {\gamma}^2 + \frac{\tau_s p}{\tau_x}) \langle s \rangle}{\langle y \rangle}$ 
& $1+0.5 \frac{{\beta}^2 \langle x \rangle}{\langle y \rangle}
+\frac{\tau_s ({\gamma}^2 + 0.5 p) \langle s \rangle}{\tau_y \langle y \rangle}$
& $1+\frac{{\beta}^2 \tau_x \langle x \rangle}{\tau_y \langle y \rangle}
+\frac{\tau_s ({\gamma}^2 + p) \langle s \rangle}{\tau_y \langle y \rangle}$ \\
&$\sigma_{sy}^2$ & $\rho(\langle y \rangle - \beta \langle x \rangle)$ 
& $ \frac{\tau_s (\langle y \rangle- \beta \langle x \rangle)}{\tau_y}$
& $\frac{\tau_s (\langle y \rangle- \beta \langle x \rangle)}{\tau_y}$ \\
\end{tabular}
\end{ruledtabular}
\end{table*}

\noindent 
While writing the above equations, we do not explicitly show the dynamical 
equation for S component but use the previously written Eq.~(\ref{eq7}) for 
the OSC. Using usual method of solution, we evaluate the following 
expression of variance and co-variance
\begin{eqnarray}
\sigma_y^2 &=& \langle y \rangle
+ \frac{\tau_y^{-1} \beta^2 \langle x \rangle
}{(\tau_x^{-1}+\tau_y^{-1})}
+ \frac{\tau_y^{-1} {\gamma}^2 \langle s \rangle
}{(\tau_s^{-1}+\tau_y^{-1})} \nonumber\\
&& +p  \frac{\tau_x^{-1} \tau_y^{-1} (\tau_s^{-1}+\tau_x^{-1}+\tau_y^{-1}) \langle s \rangle
}{(\tau_s^{-1}+\tau_x^{-1})(\tau_x^{-1}+\tau_y^{-1})(\tau_s^{-1}+\tau_y^{-1})}, \nonumber\\
\sigma_{sy}^2 &=& 
\frac{\tau_y^{-1} \langle y \rangle}{(\tau_s^{-1}+\tau_y^{-1})} 
- \frac{\tau_s^{-1} \tau_y^{-1} \beta \langle x \rangle}{(\tau_s^{-1}
+\tau_x^{-1})(\tau_s^{-1}+\tau_y^{-1})} ,
\label{eq15}
\end{eqnarray}

\noindent
where $\alpha$, $\beta$, $\gamma$ and $p$ are $k_2/\tau_x^{-1}, k_3/\tau_y^{-1}$, 
$k_3^{\prime}/\tau_y^{-1}$ and ${\alpha}^2 {\beta}^2+2\alpha \beta \gamma$,
respectively. All the rate constants define their usual meaning in the above 
kinetic equations (Eqs.~(\ref{eq13}-\ref{eq14})). To understand the effect of
relaxation time scale on this model, we take all possible relations among the 
three relaxation time constants and obtain nine conditions using which 
modified expressions of Fano factor and co-variance are calculated 
(see Table~\ref{table3}).

\begin{figure}[!b]
\begin{center}
\includegraphics[width=0.75\columnwidth,angle=0]{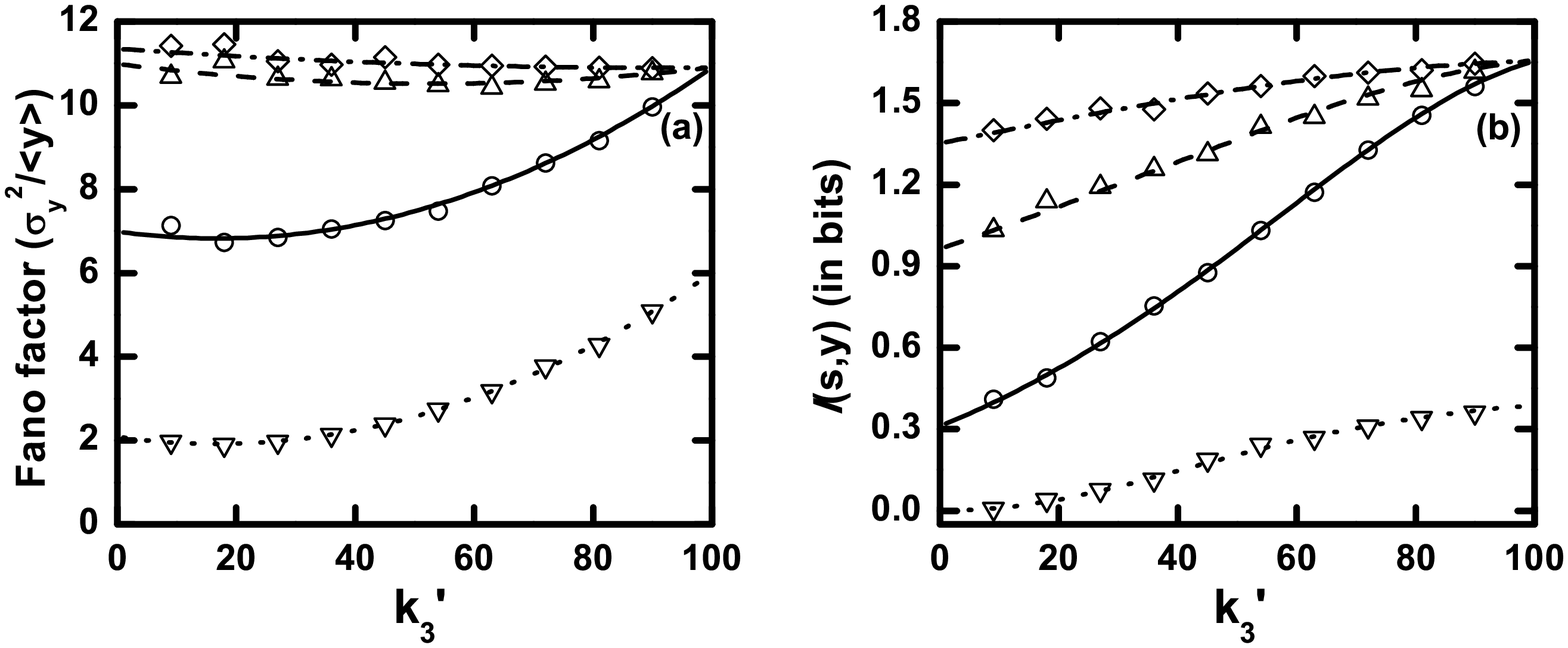}
\end{center}
\caption{The OCFFL. 
(a) Fano factor and (b) mutual information ${\cal I} (s,y)$ profiles 
as a function of S dependent synthesis rate constant $k_{3}^{'}$ 
of Y component. 
The relations $k_1/\tau_{s}^{-1} = k_2/\tau_{x}^{-1} = 10$, 
$(10 k_3+k_{3}^{'})/\tau_{y}^{-1}=10$ and $\tau_{y}^{-1}$=10 are 
maintained so that steady state population of all the components 
remain unaltered. 
In both plots, for solid (with open circles), dashed (with open upward 
triangle), dotted (with open downward triangle) and dash dotted (with 
open diamond) lines we have used
$\tau_{s}^{-1} = \tau_{x}^{-1} = 0.1$, 
$\tau_{s}^{-1} = \tau_{x}^{-1}/10 = 0.1$, 
$\tau_{s}^{-1}/100 = \tau_{x}^{-1} = 0.1$ and 
$\tau_{s}^{-1} = \tau_{x}^{-1}/100 = 0.1$, respectively. 
The symbols are generated using stochastic simulation algorithm 
\cite{Gillespie1976,Gillespie1977} and the lines are due to theoretical 
calculation.
}
\label{fig4}
\end{figure}

Similar to our previously discussed motifs, the OCFFL motif also accomplishes 
maximum, intermediate and minimum values of Fano factor and co-variance 
for the three separate relaxation time limiting conditions
$\tau_s \gg \tau_x \gg \tau_y$, $\tau_s \approx \tau_x \approx \tau_y$ and 
$\tau_s \ll \tau_x \ll \tau_y$. These modified (maximum, intermediate and 
minimum) forms are given in Table~\ref{table3}. For OCFFL, we do not explore
graphically the role of relaxation rate constant $\tau_y^{-1}$ on Fano factor
and mutual information. Due to the presence of an interesting feature in this
motif by the virtue of direct and indirect contribution of two regulatory pathways
in gene regulation via S and X, respectively, we look at the dependence of
the rate constants $k_3^{\prime}$ and $k_3$. Since, on the basis of the dominating 
power amid the two rate constants, the OCFFL motif can reduce to either OSC or 
TSC, we investigate the effect of these synthesis rate constants on the fluctuations 
and mutual information propagation. We show Fano factor and mutual information 
${\cal I} (s,y)$ as a function of $k_3^{\prime}$ in Fig.~(\ref{fig4}). For both plots, we 
maintain a constant pool of the steady state population level of the Y component 
using the relation $(10k_3 + k_3^{\prime}) / \tau_{y}^{-1}=10$ and set a high value 
of the relaxation rate constant $\tau_{y}^{-1}=10$. Only in this relaxation time 
domain, the output component can track fluctuations in the upstream signal 
very accurately.


In Fig.~(\ref{fig4}), we increase the influence of direct regulatory pathway 
with the help of augmentation of the S dependent synthesis rate constant 
$k_3^{\prime}$ of the Y component. By doing this, we increase the 
contribution of the OSC motif in the OCFFL but decrease the contribution
of X dependent synthesis rate constant $k_3$ due to the relation 
$(10k_3 + k_3^{\prime}) / \tau_{y}^{-1}=10$.
In Fig.~\ref{fig4}(a), Fano factor profile slowly goes down with $k_3^{\prime}$
for $\tau_s^{-1}=\tau_x^{-1}/10=0.1$ and $\tau_s^{-1}=\tau_x^{-1}/100=0.1$. 
At these two limits, fluctuations can propagate through this motif very smoothly. 
For very low value of $k_3^{\prime}$, the OCFFL motif behaves as a TSC 
with high Fano factor value. On the other hand, for high value of $k_3^{\prime}$,
the OCFFL motif behaves as a OSC with comparatively low Fano factor value.
For the rest of the two parameter sets ($\tau_s^{-1}=\tau_x^{-1}=0.1$ and 
$\tau_s^{-1}/100=\tau_x^{-1}=0.1$), the Fano factor value increases with 
$k_3^{\prime}$ as fluctuations propagation is hindered by the 
intermediate component X due to the indirect pathway. At these limits, for
low value of $k_3^{\prime}$, the motif attains a lower Fano factor value 
but as the value of $k_3^{\prime}$ increases, the extent of OSC character
plays a dominant role which in turn increases the Fano factor value and
fluctuations propagation becomes smooth via the direct pathway without 
any sort of intermediate obstacle.

In Fig.~\ref{fig4}(b), we show the mutual information ${\cal I} (s,y)$ in 
between the input signal S and the output Y as a function of 
$k_3^{\prime}$. The profile shows an increasing trend and is valid for 
all sets of parameter considered. Information processing is mainly 
affected by fluctuations and number of intermediate component(s) in 
between the input and the output of the corresponding network.
Hence, at low value of $k_3^{\prime}$, lesser amount of signal is
transmitted compared to high value of $k_3^{\prime}$ due to the
transition from effective TSC character to effective OSC character as 
$k_3^{\prime}$ is increased. Among the four parameter sets considered, 
exceptionally low ${\cal I} (s,y)$ value is obtained for 
$\tau_s^{-1}/100=\tau_x^{-1}=0.1$ due to faster fluctuations rate of S 
component than X and Y components. Living systems having the
OCFFL motif have a great advantage of adopting either of the two
linear cascades (OSC or TSC) with the variation of weightage on direct
or indirect pathway of gene regulation. Thus, any system when gives
an extra importance on the TSC, attains maximum output fluctuations
with minimum mutual information but attains reverse results by giving
importance to the OSC motif. A trade off between output fluctuations 
and mutual information may be accomplished by the system using 
these two pathways, direct and indirect.  Therefore, the essence of this 
motif is that during evolution it has been designed in a way that 
makes a living system more adaptable within any diverse environmental 
situation.

\subsubsection{AND like gate}

In the AND like CFFL (ACFFL) motif (Fig.~\ref{fig1}(e)), both S and X 
jointly regulate the target gene Y. Thus, in the dynamical equation of 
Y, the synthesis term is expressed in terms of both S and X. Other than
this synthesis part, rest of the equations of all the dynamical components 
are the same as the OCFFL motif,
\begin{eqnarray}
\frac{dx}{dt} & = & k_2 s - \tau_x^{-1} x + \xi_x (t), 
\label{eq16} \\
\frac{dy}{dt} & = & k_3 s x - \tau_y^{-1} y + \xi_y (t).
\label{eq17}
\end{eqnarray}


\begin{table*}[!t]
\caption{\label{table4} 
Modified forms of the analytical solution (Eq.~18) of ACFFL motif. 
Fano factor ($\sigma_{y}^2/\langle y \rangle$) and co-variance 
($\sigma_{sy}^2$) at different relaxation time limits are shown
where $\rho=\tau_s/(\tau_s+\tau_y) \leqslant 1$.
}
\begin{ruledtabular}
\begin{tabular}{ccccc}
& &$\tau_x \gg \tau_y$ & $\tau_x \approx \tau_y$ & $\tau_x \ll \tau_y$ \\
\hline
$\tau_s \gg \tau_x$ & Fano factor & $1+\frac{\langle y \rangle}{\langle x \rangle}
+4 \frac{\langle y \rangle}{\langle s \rangle}$ 
& $1+0.5 \frac{\langle y \rangle}{\langle x \rangle}
+4 \frac{\langle y \rangle}{\langle s \rangle}$
& $1+\frac{\tau_x \langle y \rangle}{\tau_y \langle x \rangle}
+4 \rho \frac{\langle y \rangle}{\langle s \rangle}$ \\
&$\sigma_{sy}^2$ & $2 \langle y \rangle$ & $ 2 \langle y \rangle$
& $2 \rho \langle y \rangle$ \\
$\tau_s \approx \tau_x$ & Fano factor & $1+\frac{\langle y \rangle}{\langle x \rangle}
+2.5\frac{\langle y \rangle}{\langle s \rangle}$ 
& $1+0.5 \frac{\langle y \rangle}{\langle x \rangle}
+\frac{13 \langle y \rangle}{ 8 \langle s \rangle}$
& $1+\frac{\tau_x \langle y \rangle}{\tau_y \langle x \rangle}
+4 \frac{\tau_x \langle y \rangle}{\tau_y \langle s \rangle}$ \\
&$\sigma_{sy}^2$ & $1.5 \langle y \rangle$ & $ 0.75 \langle y \rangle$
& $1.5 \frac{\tau_s \langle y \rangle}{\tau_y}$ \\
$\tau_s \ll \tau_x$ & Fano factor & $1+ \frac{\langle y \rangle}{\langle x \rangle} 
 + \frac{(\rho + 3 \frac{\tau_s}{\tau_x})\langle y \rangle}{\langle x \rangle}$ 
& $1+0.5 \frac{\langle y \rangle}{\langle x \rangle}
+2.5 \frac{\tau_s \langle y \rangle}{\tau_x \langle s \rangle}$
& $1+\frac{\tau_x \langle y \rangle}{\tau_y \langle x \rangle}
+4\frac{\tau_s \langle y \rangle}{\tau_y \langle s \rangle}$ \\
&$\sigma_{sy}^2$ & $ (1 + \frac{\tau_s}{\tau_x}) \rho \langle y \rangle$ 
& $ (1 + \frac{\tau_s}{\tau_x}) \frac{\tau_s \langle y \rangle}{\tau_y}$
& $(1 + \frac{\tau_s}{\tau_x}) \frac{\tau_s \langle y \rangle}{\tau_y}$ \\
\end{tabular}
\end{ruledtabular}
\end{table*}

\noindent
In the above set of equations, all the rate constants define kinetic significance 
of the corresponding network components. Here, we also do not rewrite 
the dynamical equation for S and use the previous equation (Eq.~(\ref{eq7})).
We solve the set of dynamical equations considering an approximation 
$\langle sx \rangle = \langle s \rangle \langle x \rangle$ 
\cite{Elf2003,Swain2004,Bialek2005, Kampen2005, deRonde2010}
as reactions in between the two components S and X are uncorrelated 
with each other and obtain simplified analytical form of variance and
co-variance of the target gene Y as
\begin{eqnarray}
\sigma_y^2 &=& \langle y \rangle
+ \frac{\tau_y^{-1} \langle y \rangle^2
}{(\tau_x^{-1}+\tau_y^{-1}) \langle x \rangle}
+ \frac{\tau_y^{-1} \langle y \rangle^2
}{(\tau_s^{-1}+\tau_y^{-1}) \langle s \rangle} \nonumber\\
&&+ \frac{3 \tau_x^{-1} \tau_y^{-1} (\tau_s^{-1}+\tau_x^{-1}+\tau_y^{-1}) \langle y \rangle^2
}{(\tau_s^{-1}+\tau_x^{-1})(\tau_x^{-1}+\tau_y^{-1})(\tau_s^{-1}+\tau_y^{-1}) \langle s \rangle}, 
\nonumber\\
\sigma_{sy}^2 &=& 
\frac{\tau_y^{-1} \langle y \rangle}{(\tau_s^{-1}+\tau_y^{-1})} 
+ \frac{\tau_x^{-1} \tau_y^{-1} \langle y \rangle}{(\tau_s^{-1}+\tau_x^{-1})(\tau_s^{-1}+\tau_y^{-1})} .
\label{eq18}
\end{eqnarray}

It is interesting to note that for both TSC and ACFFL, we get almost 
equivalent variance expression (see Eq.~(\ref{eq12}) and Eq.~(\ref{eq18})) 
except for two factors. These extra terms are the third and the
fourth term on the right hand side of Eq.~(\ref{eq18}). The third term
arises due to direct regulation of Y by S. The fourth term is multiplied
by a numerical factor 3. From these extra terms, it is obvious that ACFFL
shows higher fluctuating property than TSC due to the additive nature
of two positive regulatory pathways (both direct and indirect)
\cite{Kaern2005, Alon2007, Raj2008, Eldar2010, SilvaRocha2010}. 
From Eq.~(\ref{eq18}), we get all possible reduced forms of Fano factor 
and co-variance expressions using nine possible relations among the 
three relaxation time scales (see Table~\ref{table4}).

\begin{figure}[!b]
\begin{center}
\includegraphics[width=0.75\columnwidth,angle=0]{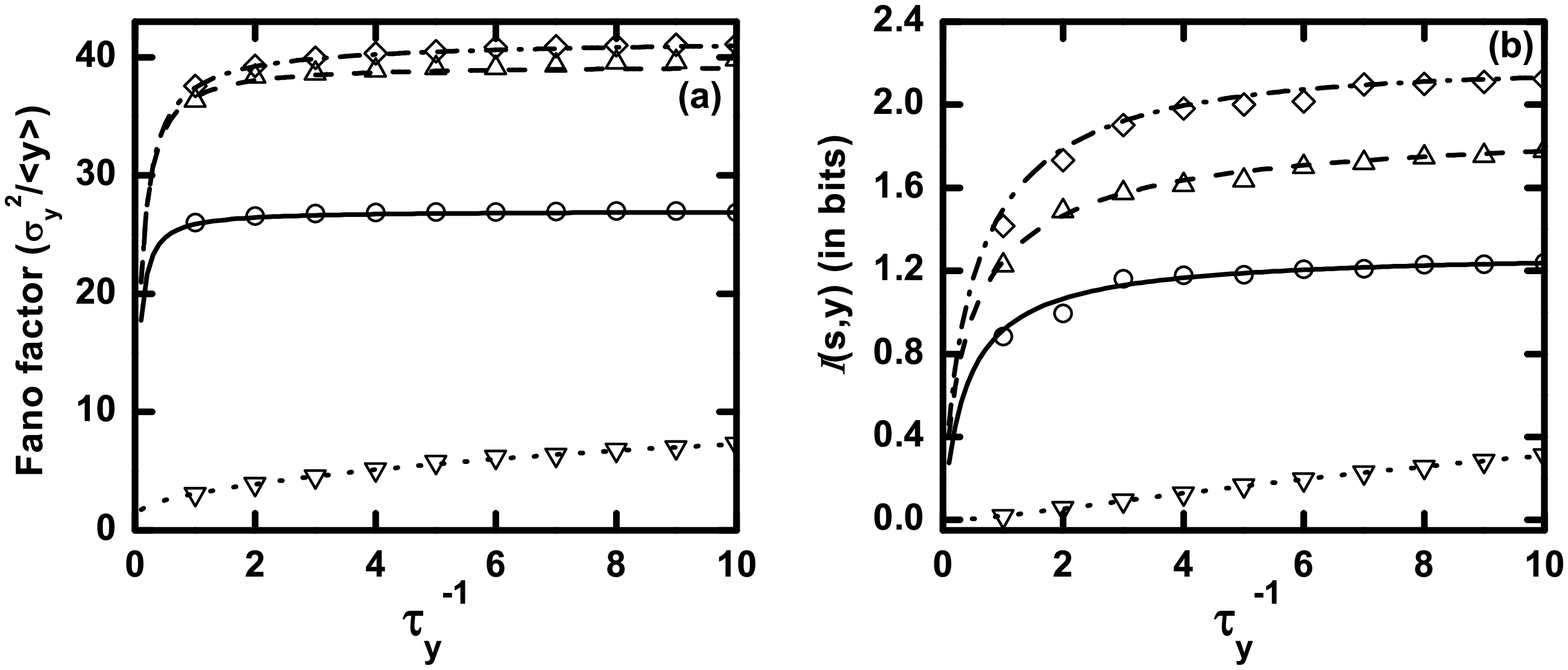}
\end{center}
\caption{The ACFFL. 
(a) Fano factor and (b) mutual information ${\cal I} (s,y)$ profiles as a
function of relaxation rate constant $\tau_y^{-1}$ of Y component.
The relations $k_1/\tau_{s}^{-1} = k_2/\tau_{x}^{-1} = 10$ and 
$k_3/\tau_{y}^{-1} = 0.1$ are maintained so that steady state population 
of all the components remain unaltered. 
In both plots, for solid (with open circles), dashed (with open upward triangle), 
doted (with open downward triangle) and dash dotted (with open diamond) 
lines we have used
$\tau_{s}^{-1} = \tau_{x}^{-1} = 0.1$, 
$\tau_{s}^{-1} = \tau_{x}^{-1}/10 = 0.1$, 
$\tau_{s}^{-1}/100 = \tau_{x}^{-1} = 0.1$ and 
$\tau_{s}^{-1} = \tau_{x}^{-1}/100 = 0.1$, 
respectively. 
The symbols are generated using stochastic simulation algorithm 
\cite{Gillespie1976,Gillespie1977} and the lines are due to theoretical 
calculation.}
\label{fig5}
\end{figure}

As shown in the calculation for previous motifs, it is clear from Table~\ref{table4} 
that, for ACFFL, Fano factor and co-variance achieve maximum, intermediate 
and minimum values for $\tau_s \gg \tau_x \gg \tau_y$, 
$\tau_s \approx \tau_x \approx \tau_y$ and $\tau_s \ll \tau_x \ll \tau_y$, 
respectively. At these time scales, the modified forms of both Fano factor 
and co-variance are almost similar with the modified forms of TSC (see 
Table~\ref{table2}) but terms like $\langle y \rangle/\langle s \rangle$ and 
$\langle y \rangle$ appear with multiplicative factor greater than 1. This
leads to a high level of fluctuating environment for the ACFFL motif.
The main reason behind the elevation of output fluctuations is the addition 
of fluctuations due to input signal S into the total fluctuations of the output 
component Y in two ways, direct and indirect pathways. Due to the such 
types of  fluctuations addition phenomena, we get a higher Fano factor value. 
Similarly, we also obtain high level of mutual information ${\cal I} (s,y)$ 
transduction due to the presence of two subsequent pathways by which 
the target gene reliably accumulates signal information with greater extent 
and transcribes gene products precisely with the variation of input signal. 
To verify these features, we show Fano factor and mutual information as a
function of relaxation rate constant $\tau_y^{-1}$ for four different parameter 
sets of relaxation rate constants in Fig.~\ref{fig5}.


In Fig.~\ref{fig5}(a), highest Fano factor value is obtained for the parameter
sets $\tau_s^{-1}=\tau_x^{-1}/10=0.1$ and $\tau_s^{-1}=\tau_x^{-1}/100=0.1$.
Both sets are due to faster fluctuations of the intermediate component  X. On the
other hand, lowest Fano factor value is attained for $\tau_s^{-1}/100=\tau_x^{-1}=0.1$ 
and is due to faster fluctuations of input signal S. For $\tau_s^{-1}=\tau_x^{-1}=0.1$, 
the motif shows an intermediate Fano factor value. In all cases, Fano factor value 
increases with the relaxation rate constant of the target gene Y. Similar trend is 
also observed for the mutual information ${\cal I} (s,y)$ profile (Fig.~\ref{fig5}(b)). 
Therefore, this motif has high information processing capacity in spite of the
presence of high level of fluctuations. Such high input signal processing 
phenomena facilitate the networkÕs reliability for signal transduction in GTRN 
and makes this motif a highly abundant one among rest of the CFFL present
within the family of cellular networks. In this connection, it is important to mention 
that the network architecture can facilitate cellular fitness advantage in adverse 
environment due to increase of fluctuations. Thus, high fluctuations in output 
gene expression trigger phenotypic heterogeneity in clonal cell populations and 
can induce drug resistance \cite{Charlebois2011,Charlebois2014}. A similar type 
of CFFL is also liable for drug resistance of human cancer cells.

\subsection{Incoherent feed forward loop}

The last motif considered in the present work is another class of FFL known 
as Incoherent feed forward loop. We focus only on the type-1 incoherent feed 
forward loop (ICFFL) (Fig.~\ref{fig1}(f)). In this motif, two regulatory pathways act in 
an opposite manner. Here, input signal S positively regulates the target gene Y
through the direct pathway. However, the intermediate component X represses
the expression of Y. As a result, S initially activates both X and Y rapidly but after 
some time, population level of X reaches a threshold to repress the production of 
Y. While modeling the repression phenomenon, we consider the Hill coefficient to
be one. Thus, the Langevin equations for the dynamical quantities can be written 
as
\begin{eqnarray}
\frac{dx}{dt} & = & k_2 s - \tau_x^{-1} x + \xi_x (t), 
\label{eq19} \\
\frac{dy}{dt} & = & k_3 g(x)s - \tau_y^{-1} y + \xi_y (t).
\label{eq20}
\end{eqnarray}


\begin{table*}[!t]
\caption{\label{table5} 
Modified form of the analytical solution (Eq.~(\ref{eq21})) of ICFFL motif.
Fano factor ($\sigma_{y}^2/\langle y \rangle$) and co-variance
($\sigma_{sy}^2$) at different relaxation time limits are shown 
where $\rho=\tau_s/(\tau_s+\tau_y) \leqslant 1$.
}
\begin{ruledtabular}
\begin{tabular}{ccccc}
& &$\tau_x \gg \tau_y$ & $\tau_x \approx \tau_y$ & $\tau_x \ll \tau_y$ \\
\hline
$\tau_s \gg \tau_x$ & Fano factor & $1+\frac{\langle y \rangle}{\langle x \rangle}$ 
& $1+0.5 \frac{\langle y \rangle}{\langle x \rangle}$
& $1+\frac{\tau_x \langle y \rangle}{\tau_y \langle x \rangle}$ \\
&$\sigma_{sy}^2$ & $0$ & $0$& $0$ \\
$\tau_s \approx \tau_x$ & Fano factor & $1+\frac{\langle y \rangle}{\langle x \rangle}
+0.5\frac{\langle y \rangle}{\langle s \rangle}$ 
& $1+0.5 \frac{\langle y \rangle}{\langle x \rangle}
+\frac{\langle y \rangle}{8 \langle s \rangle}$
& $1+\frac{\tau_x \langle y \rangle}{\tau_y \langle x \rangle}$ \\
&$\sigma_{sy}^2$ & $0.5 \langle y \rangle$ & $ 0.25 \langle y \rangle$
& $0.5 \frac{\tau_s \langle y \rangle}{\tau_y}$ \\
$\tau_s \ll \tau_x$ & Fano factor & $1+\frac{\langle y \rangle}{\langle x \rangle}
+\frac{(\rho - \frac{\tau_s}{\tau_x}) \langle y \rangle}{\langle s \rangle}$ 
& $1+0.5 \frac{\langle y \rangle}{\langle x \rangle}
+0.5 \frac{\tau_s \langle y \rangle}{\tau_y \langle s \rangle}$
& $1+\frac{\tau_x \langle y \rangle}{\tau_y \langle x \rangle}$ \\
&$\sigma_{sy}^2$ & $(1 - \frac{\tau_s}{\tau_x}) \rho \langle y \rangle$ & 
$ (1 - \frac{\tau_s}{\tau_x}) \frac{\tau_s \langle y \rangle}{\tau_y}$
& $(1 - \frac{\tau_s}{\tau_x}) \frac{\tau_s \langle y \rangle}{\tau_y}$ \\
\end{tabular}
\end{ruledtabular}
\end{table*}

\noindent
In the above equation, $g(x)=K/(K+x)$ is a nonlinear repressive function that
depends on the concentration of X and $K$, the ratio of unbinding to binding 
rate constants of transcription factor X at the promoter region of Y gene. We 
also consider $\langle g(x)s \rangle=\langle g(x) \rangle \langle s \rangle$ as 
the kinetic equations for both  S and X are uncorrelated with each other and 
$\langle x \rangle/(K + \langle x \rangle) \approx1$ as steady state concentration 
of the X component is much higher than the unbinding-binding constant ($x \gg K$).
For this motif, we use the earlier kinetic equation for S (see Eq.~(\ref{eq7})). Solving
the set of kinetic equations, we get the simplified mathematical form of variance and 
co-variance \cite{Elf2003, Swain2004, Bialek2005, Kampen2005, deRonde2010}
\begin{eqnarray}
\sigma_y^2 &=& \langle y \rangle
+ \frac{\tau_y^{-1} \langle y \rangle^2
}{(\tau_x^{-1}+\tau_y^{-1})\langle x \rangle} 
+ \frac{\tau_y^{-1} \langle y \rangle^2
}{(\tau_s^{-1}+\tau_y^{-1})\langle s \rangle}\nonumber\\
&& - \frac{\tau_x^{-1}\tau_y^{-1} (\tau_s^{-1}+\tau_x^{-1}+\tau_y^{-1}) \langle y \rangle^2
}{(\tau_s^{-1}+\tau_x^{-1})(\tau_x^{-1}+\tau_y^{-1})(\tau_s^{-1}+\tau_y^{-1})\langle s \rangle}, \nonumber\\
\sigma_{sy}^2 &=& 
\frac{\tau_y^{-1} \langle y \rangle
}{(\tau_s^{-1}+\tau_y^{-1})} 
- \frac{\tau_x^{-1}\tau_y^{-1} \langle y \rangle}{(\tau_s^{-1}+\tau_x^{-1})(\tau_s^{-1}+\tau_y^{-1})} .
\label{eq21}
\end{eqnarray}


\begin{figure}[!t]
\begin{center}
\includegraphics[width=0.75\columnwidth,angle=0]{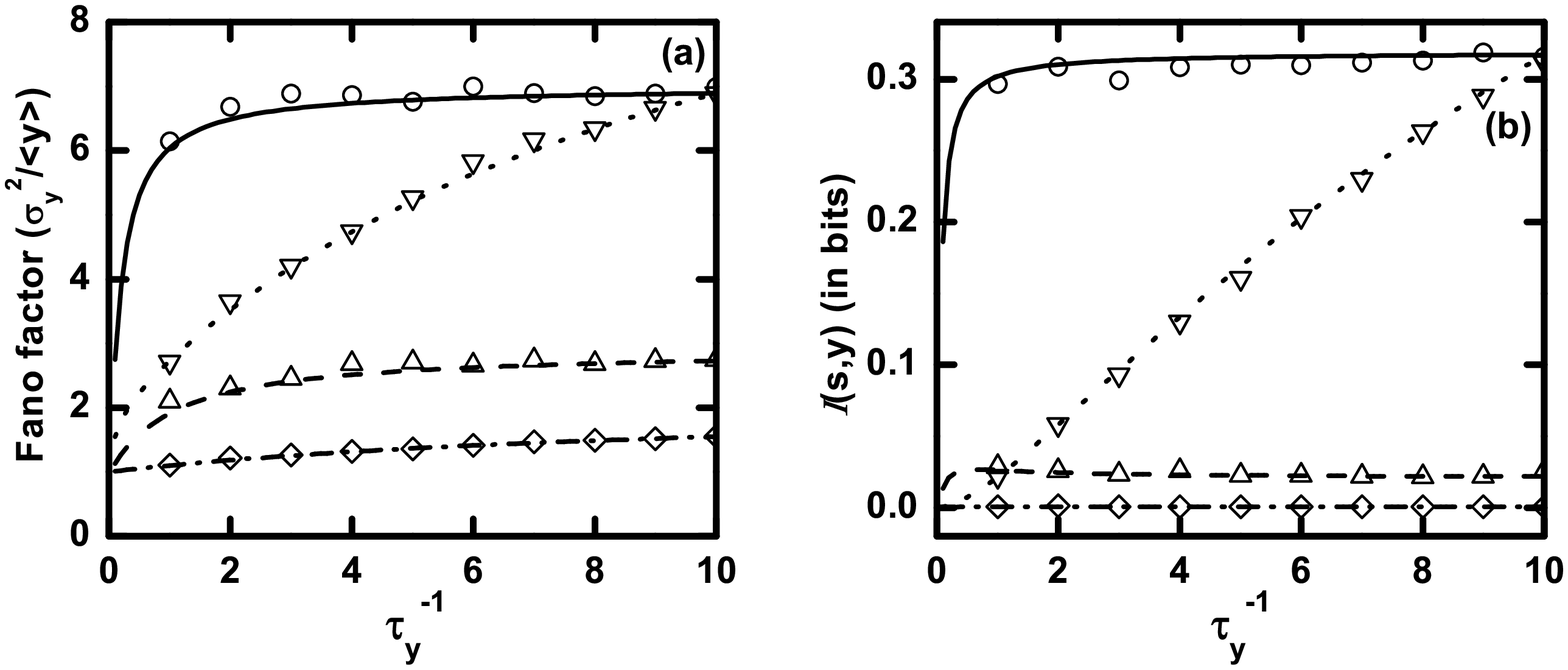}
\end{center}
\caption{The ICFFL. (a) Fano factor and (b) mutual information ${\cal I} (s,y)$ 
profiles as function of relaxation rate constant $\tau_{y}^{-1}$ of Y component. 
$k_1/\tau_{s}^{-1} = k_2/\tau_{x}^{-1} = 10$ and $k_3/\tau_{y}^{-1} = 1010$, 
ratios are maintained throughout these plots, so steady state population of 
all the components remain unaltered. In both plots, for 
solid (with open circles), dash (with open upward triangle), 
dot (with open downward triangle) and dash dot (with open diamond) 
lines, we use the following sets of 
parameters: $\tau_{s}^{-1} = \tau_{x}^{-1} = 0.1$, $\tau_{s}^{-1} = \tau_{x}^{-1}/10 = 0.1$, 
$\tau_{s}^{-1}/100 = \tau_{x}^{-1} = 0.1$ and $\tau_{s}^{-1} = \tau_{x}^{-1}/100 = 0.1$, 
respectively. The symbols are generated using stochastic simulation algorithm and 
the lines are from theoretical calculation.The symbols are generated using stochastic 
simulation algorithm \cite{Gillespie1976,Gillespie1977} and the lines are due to theoretical 
calculation.
}
\label{fig6}
\end{figure}

\noindent
In the above variance expression, two S dependent fluctuations terms are 
present with opposite sign thus compensating each other. The variance
expression is quite similar to the expression of TSC (see Eq.~(\ref{eq12})) 
but differs in the sign of the last term and the appearance of an extra term due to
direct regulation of Y by S. If one calculates the magnitude of variance for
both TSC and ICFFL using unique steady state population level of all three
components and the corresponding relaxation time scale, a higher value
of $\sigma_y^2$ will be observed for TSC compared to ICFFL. To check the 
validity of this effect, we consider all possible relations among the three 
relaxation times of the corresponding network components and get nine 
possible relations. Using these relations, modified forms of Fano factor and 
co-variance are calculated and are given in Table~\ref{table5}. From these
modified expressions, it is clear that ICFFL exerts lesser amount of fluctuations
compared to the other motifs considered in this work while developing the 
target gene product Y. Development of low fluctuations thus gets reflected
in the Fano factor value. This happens due to repression of target gene by
X which effectively reduces fluctuations associated with Y
\cite{Kaern2005, Alon2007, Raj2008, Eldar2010, SilvaRocha2010}. 
It is important to note that, in Table~\ref{table5}, under some relaxation time
scale limits, the Fano factor expressions are free from terms with
$\langle s \rangle$, the S dependent fluctuations. This happens due to a
simultaneous contribution of direct and indirect pathways which cancel 
each other by having equal magnitude but opposite sign. As a consequence, 
some of the co-variance values also become zero. This suggests that at
these time scale limits, the system cannot incorporate the information of the
input signal properly, thereby transducing information about the input signal
unreliably to the target gene. Keeping this in mind, we explore the nature
of  Fano factor and mutual information ${\cal I} (s,y)$ as a function of
relaxation rate constant $\tau_y^{-1}$ of Y component using four different
sets, i.e., $\tau_s^{-1}=\tau_x^{-1}=0.1$, $\tau_s^{-1}=\tau_x^{-1}/10=0.1$, 
$\tau_s^{-1}/100=\tau_x^{-1}=0.1$ and $\tau_s^{-1}=\tau_x^{-1}/100=0.1$.

In Fig.~\ref{fig6}(a), Fano factor value gradually increases with the relaxation 
rate constant $\tau_{y}^{-1}$. Very low level of Fano factor value is attained by 
the motif for $\tau_s^{-1}=\tau_x^{-1}/10=0.1$ and $\tau_s^{-1}=\tau_x^{-1}/100=0.1$. 
For these two sets, input fluctuations flow successfully through the direct and 
indirect pathways. Thus, two S dependent fluctuating terms that are equal in 
magnitude but opposite in sign compensate each other, consequently 
suppressing the fluctuations associated with the network. On the other hand,
for the other two parameter sets, a high level of Fano factor values is found. 
This happens due to slower rate of fluctuations in the intermediate X. The 
input fluctuations that come through the indirect pathway are filtered out by 
X for its low pass filter nature. As a result, two S dependent fluctuating terms 
do not completely cancel each other. In these parameter sets, the fluctuating 
part that contributes to the direct pathway shows its prominent effect than the 
fluctuations due to indirect pathway which finally gets reflected in the total 
output fluctuations. Likewise, in Fig.~\ref{fig6}(b), mutual information 
${\cal I} (s,y)$ values are near to zero for the first two parameter sets giving 
low level of Fano factor and significant ${\cal I} (s,y)$ values for the rest of 
the parameter sets.


\section{Conlusion}

In this paper, we have analyzed signal transmission of a fluctuating 
input signal through highly specialized biochemical signaling networks. 
We have analytically calculated Fano factor of output and mutual 
information between input and output signals for two linear and three 
branched cascades to comprehend the significance of these networks 
in biological systems. On the basis of linear noise approximation, we 
have solved nonlinear chemical Langevin equations and verified the
results with exact stochastic simulation of the corresponding nonlinear 
networks and found that the approximation method is quite accurate. 
In the analytical calculation, we have considered that all noise terms 
($\xi_s, \xi_x, \xi_y$) are Gaussian in nature and the effect of cross
correlation between two noise terms is zero. We have calculated Fano 
factor and mutual information for five networks with the variation of 
relaxation rate constants of all network components and have studied 
effect of input signal on these two measurable quantities. Our study 
not only takes care of individual motifs but also presents a comparative 
study of all the network motifs while considering Fano factor and mutual 
information. For graphical presentation of the output component of each
motif, we have tuned synthesis and relaxation rate constant of each 
network components in such a way so that the steady state population 
of the network components remains constant as well as the total population 
of any network is also preserved. Adoption of such a policy helps us to
apprehend how Fano factor and mutual information values get affected
from one motif to another under equal population of network components.

We have started our calculations considering linear type of cascades and 
the first motif we have considered is OSC where the motif can precisely 
characterize the information of input signal at faster relaxation time of the
output component compared to the input one. This accuracy gradually 
decreases with the increment of input relaxation rate. We have also shown 
that the OSC motif is unable to differentiate the variation of input signal at 
high population limit of input component. We then compare this motif with 
the standard gene regulation network to explicate such time scale effect on 
gene regulation and found some significant circumstances in which the
gene regulatory network can tune up optimum fluctuations for both 
essential and nonessential proteins. Our analysis is at par with the results 
of Fraser et al \cite{Fraser2004} where they have performed analysis 
using several experimentally determined gene regulation rates.

The second motif TSC is then considered by introducing an intermediate 
component in OSC. Through our analysis, we have observed that Fano 
factor of output gets amplified in magnitude. In addition, similar kind of 
relaxation time scale effect for propagation of fluctuations as observed
in OSC, is also present in this motif. Our analysis suggests that relaxation
time scale of intermediate component is a crucial factor for signal transmission
through this motif and can control fluctuations associated with the output.
The intermediate component  acts as a low pass filter for very fast fluctuations 
and hinders input fluctuations that flow through the TSC motif. After analyzing
the TSC, we have introduced $n$ number of intermediate components in 
between the input and the output component to generate a generalized 
linear long chain cascade and derived simplified form of Fano factor 
expression at three distinct time scale limits. We have shown that the 
output fluctuations increase with number of intermediate components 
and this result agrees with previously published several experimental 
and theoretical results. The main utility of these expressions is that one 
can easily calculate fluctuations associated with any linear long chain
cascade without enough knowledge of the parameter values of the 
network.

Next, we have chosen FFL, a group of branched pathways that are 
abundant in the signal transduction machinery of living systems and 
are generated by lateral combination of OSC and TSC with different 
modes of interaction. At first, we have studied OCFFL from FFL group 
and calculated Fano factor and mutual information with the
extent of direct (OSC) and indirect (TSC) contribution of input signal
to the target gene. In our calculation of Fano factor, we have observed
two opposing behavior while tuning the control parameter. In one case,
Fano factor decreases slowly while in the other situation, it shows an
increasing trend. These results together suggest that OCFFL performs
differently for diverse relaxation time scale limits. Furthermore, depending
on the contribution of the direct and the indirect pathway, it can acquire
the character of OSC and TCF motif, respectively. Such quality of OCFFL 
helps a living system to survive in diverse environmental conditions. We
then extend our analysis to ACFFL where a high value of Fano factor and
mutual information is observed. Such high values of Fano factor and 
mutual information reveals that ACFFL motif can transduce the signal with
high reliability and supports its ubiquitous presence in several biological
species. The last motif we have considered is ICFFL where output fluctuations 
are suppressed as the target gene is simultaneously regulated negatively 
(via indirect pathway with TSC character) as well as positively (via direct 
pathway with OSC character) by the input signal. This leads to a very low
value of Fano factor and mutual information for ICFFL motif. Such low value
suggests that living system containing this motif faces minimum fluctuations
with low mutual information propagation.

To conclude, we emphasize that our methodology is very much general 
and is applicable for other network motifs under single cell environment. 
At this point, it is important to mention that enough single cell data are not 
available for FFL. Our theoretical results thus can act as a starting point 
to verify the stochastic nature of the network motifs we have considered
in the present work. The nature of the results we have predicted can be
tested by performing experiment in a single cell scenario. We expect that 
our work will influence several experimentalists in the coming days to
understand the single cell behavior of these network motifs.


\begin{acknowledgments}
We thank Debi Banerjee for critical reading of the manuscript.
AKM and PC are thankful to University Grants Commission, 
New Delhi, for research fellowship (UGC/776/JRF(Sc)) and for 
a major research project, respectively.
SKB acknowledges financial support from CSIR, India
[01(2771)/14/EMR-II] and Bose Institute (through Institutional 
Programme VI - Development of Systems Biology).
\end{acknowledgments}


\end{document}